\documentclass[twocolumn,showpacs,preprintnumbers,amsmath,amssymb]{revtex4}

\usepackage{graphicx}
\usepackage{dcolumn}
\usepackage{bm}

\begin{document}

\preprint{ }


\title{Probing minimal scattering events in enhanced backscattering
of light \\using low-coherence induced dephasing
}

\author{Young L. Kim}
\author{Prabhakar Pradhan}
\author{Hariharan Subramanian}
\author{Yang Liu}
\author{Min H. Kim}
\author{Vadim Backman}

\affiliation{Biomedical Engineering Department, Northwestern
University,
Evanston, IL 60208}%



\begin{abstract}

We exploit low spatial coherence illumination to dephase
time-reversed partial waves outside its finite coherence area,
which virtually creates a controllable coherence volume and
isolates the minimal number of scattering events from higher order
scattering in enhanced backscattering (EBS, also known as coherent
backscattering) of light.  We report the first experimental
evidence that the minimum number of scattering events in EBS is
double scattering in discrete random media, which has been
hypothesized since the first observation of EBS of light.  We
discuss several unique characteristics and potential applications
of low-coherence EBS in weakly scattering random media.

\end{abstract}

\maketitle

Enhanced backscattering (EBS), otherwise known as coherent
backscattering, is a spectacular manifestation of
self-interference effects in elastic light scattering, which gives
rise to an enhanced scattered intensity in the backward direction.
For a plane wave illuminating a semi-infinite random medium, every
photon scattered from the medium in the backward direction has a
time-reversed photon traveling along the same path in the opposite
direction.
These photons have the same phase at
the exit points and thus interfere constructively to each other,
resulting in EBS.
Since the first observations of EBS of light in aqueous
suspensions~\cite{van_albda_Wolf_Kuga},
the EBS phenomenon has been an object of intensive investigations
in a variety of different systems such as strong scattering
materials~\cite{Wiersma:PRL:1995:Recur}, cold
atoms~\cite{Labeyrie:2003}, liquid
crystals~\cite{Sapienza:PRL:2004}, photonic
crystals~\cite{Huang:PRL:2001}, amplifying
materials~\cite{Wiersma:PRL:1995:Amp}, solar system
bodies~\cite{Mishchenko:PSS:1993}, and biological
tissues~\cite{Yoo:AO:1990,Kim:OL:2004,Kim:OL:2005}.
The dependency of the profile of EBS peaks on the path length was
also studied using time-resolved
measurements~\cite{Vreeker:PLA:1988,Yoo:AO:1990,Tourin:1997,Barabanenkov:1999}.
Moreover, the time-reversed invariance was altered using faraday
rotation generated by a strong external magnetic
field~\cite{Lenke:EPL:2000}, a phase screw
dislocation~\cite{Schwartz:OL:2005}, or the quantum internal
structure of cold atoms~\cite{Labeyrie:2003}.

\begin{figure}
\includegraphics[scale=0.87]{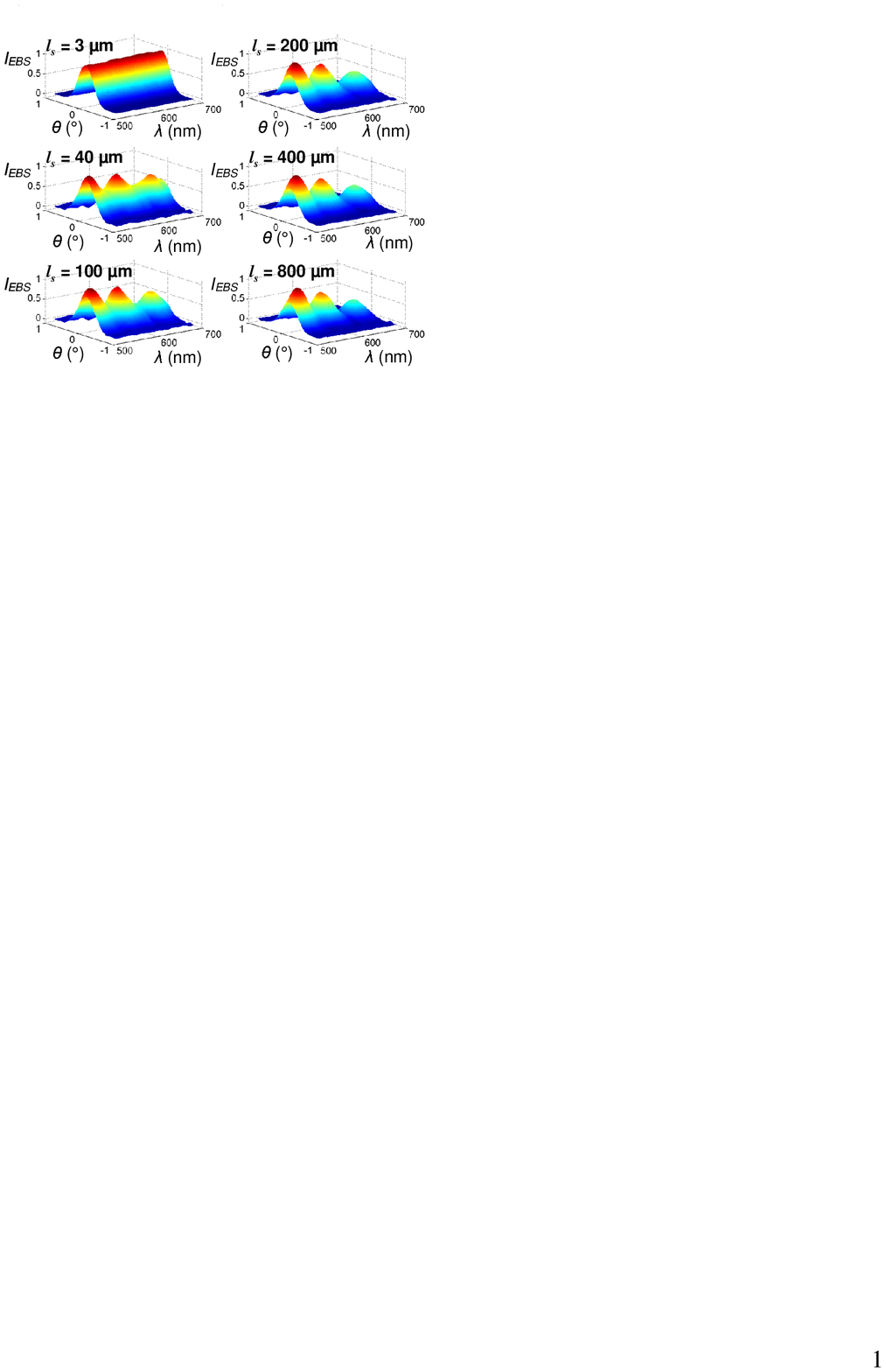}
\caption{\label{fig:fig1} EBS intensity $I_{EBS}(\theta,\lambda)$
obtained from the aqueous suspension of microspheres ($d$ = 1.5
$\mu m$) under low spatial coherence illumination ($L_{sc} =
35~\mu m$) for various scattering mean free paths $l_s$. For
$l_s\!>>\!L_{sc}$, the spectral shape remains unchanged,
indicating that it reaches to the minimal scattering events
required for EBS. }
\end{figure}



Recently, we demonstrated experimentally that low spatial
coherence illumination dephases the conjugated time-reversed paths
outside its spatial coherence area and rejects long scattering
paths resulting in a broad EBS peak (i.e., $L_{sc}\!<<l_s^*$,
where $L_{sc}$ is the spatial coherence length and $l_s^*$ is the
transport mean free path length of light in the
medium)~\cite{Kim:OL:2004,Kim:OL:2005}. (EBS under low spatial
coherence illumination is henceforth referred to as low-coherence
EBS, LEBS). LEBS possesses novel and intriguing properties:
speckle reduction and several orders of magnitude broadening of
the EBS peak, which facilitate depth-resolved measurements by
probing different scattering angles within the EBS peak. The
rationale for investigation of LEBS is further emphasized by our
demonstration that LEBS can be used to detect early precancerous
alterations in the colon far earlier than any other currently
available molecular and genetic
techniques~\cite{Kim:OL:2004,Kim:OL:2005}.







In this Letter, 
we demonstrate that dephasing induced by low spatial coherence
illumination in EBS isolates double scattering from higher order
scattering in a discrete random medium. We further show for the
first time to our knowledge the direct experimental evidence that
the minimal scattering events to generate an EBS peak in a
discrete random medium is double scattering. Our main finding is
that LEBS isolates double scattering from higher order scattering
when $L_{sc}$ is on the order of the scattering mean free path
$l_s$ of light in the medium ($l_s = l_s^*(1-g)$, where $g$ is the
anisotropy factor). From previous theoretical
studies~\cite{van_der_Mark:PRB:1988,Akkermans:JDP:1988}, it is
known that double scattering is the minimal scattering events that
are required to generate an EBS peak, because single scattering
contributes to the incoherent baseline backscattering signal but
not to the EBS peak. We take advantage of low spatial coherence
illumination to generate a finite spatial coherence area on the
sample, which in turn defines virtually a narrow elongated
coherence volume in a large volume of a weakly scattering medium
such as biological tissue.



In the search of the minimal scattering events for EBS, we
performed the followings: First, we measured spectral and angular
distributions of EBS from discrete random media consisting of the
aqueous suspensions of microspheres.
Second, we compared these spectra with the predictions of a Mie
theory-based double scattering model. Third, we validated the
angular profile of the LEBS peaks using the double scattering
model.  Finally, we investigated the polarization properties of
LEBS to further confirm the double scattering model of LEBS.




In our EBS experiments, we combined EBS measurements with low
spatial coherence, broadband illumination and spectrally-resolved
detection.  Our experimental setup was described in detail
elsewhere~\cite{Kim:OL:2004}.  In brief, a beam of broadband
cw-light from a 100 W xenon lamp (Spectra-Physics Oriel) was
collimated using a 4-$f$ lens system (divergence angle ranging
from $\sim0.04^\circ$ for $L_{sc}=200~\mu m$ to $\sim0.30^\circ$
for $L_{sc} = 35~\mu m$), polarized, and delivered onto a sample
at $\sim10^\circ$ angle of incidence to prevent the collection of
the specular reflection. By changing the aperture size in the
4-$f$ lens system, we varied spatial coherence length $L_{sc}$ of
the incident light from 200 $\mu m$ to 35 $\mu m$.  The value of
$L_{sc}$ was confirmed by the double-slit interference
experiments~\cite{Bron:1999}.
The light backscattered by the sample was collected by a sequence
of a lens, a linear analyzer (Lambda Research Optics), and an
imaging spectrograph (Acton Research). The spectrograph was
positioned in the focal plane of the lens and coupled with a CCD
camera (Princeton Instruments). The lens projected the angular
distribution of the backscattered light onto the slit of the
spectrograph. Then, the imaging spectrograph dispersed this light
according to its wavelength in the direction perpendicular to the
slit and projected it onto the CCD camera. Thus, the CCD camera
recorded a matrix of scattered intensity as a function of
wavelength~$\lambda$ and backscattering angle~$\theta$.

For spectroscopic EBS measurements, the linear analyzer was
oriented along the polarization of the incident light, which
provided a linear parallel channel.  The spectrally-resolved EBS
signals were normalized as $I_{EBS}(\theta,\lambda)=
(I(\theta,\lambda)-I_{BASE}(\lambda ))/I_{REF}(\lambda)$, where
$I(\theta,\lambda)$ is the total scattered intensity,
$I_{BASE}(\lambda)$ is the baseline (incoherent) intensity
measured at large backscattering angles ($\theta>3^\circ$), and
$I_{REF}(\lambda)$ is a reference intensity collected from a
reflectance standard (Ocean Optics).
The resulting EBS signal $I_{EBS}(\theta,\lambda)$ is referred
hereafter to as the EBS intensity.  We also investigated the
effect of the degree of circular polarization on EBS.  The degree
of circular polarization of EBS was analyzed by means of an
achromatic quarter-wavelet plate (Karl Lambrecht) positioned
between the sample and the beam splitter.
In this studies, the EBS peak was normalized by the baseline
scattering intensity.

LEBS possesses unique advantageous features compared to
conventional EBS: (i) The independent coherence area (or the
transverse modes) can be as small as a few tens of microns.  Thus,
$L_{sc}$ can be made to be the shortest length scale (except
particle sizes) in weakly scattering media such as biological
tissue (in tissue, $l_s^*$ is on the order of a few millimeters).
(ii) LEBS provides statistical information about the optical
properties of random media. A single LEBS reading averages over
multiple independent coherence areas (or channels), which reduce
the complications of realization averaging.  For example, for
$L_{sc} = 35~\mu m$, the number of independent coherence areas
$(D/L_{sc})^2 \approx 7000$, where $D = 3~mm$ is the diameter of
illumination area on the sample. (iii) LEBS allows varying
$L_{sc}$ to control the dephasing rate externally and LEBS does
not require complicated sample preparations. Thus, these
characteristics of LEBS facilitate investigations of EBS in weakly
scattering random media including biological tissue.

%



We used discrete random media consisting of aqueous suspensions of
polystyrene microspheres ($n_{sphere}$ = 1.59 and $n_{water}$ =
1.334 at $\lambda=550~nm$) (Duke Scientific) of various diameters
from 200 $nm$ to 1.5 $\mu m$. The dimension of the samples was
$\pi \times 25^2~mm^2 \times 50~mm$. We varied the scattering mean
free path $l_s$ from 3 $\mu m$ to $\sim$1000 $\mu m$ for two
selected different values of $L_{sc}$ ($L_{sc} = 110~\mu m$ and
$L_{sc} = 35~\mu m$).  The optical scattering properties of the
samples were calculated using Mie theory~\cite{van_de_Hulst:1995}.

Figure 1 shows representative EBS intensity
$I_{EBS}(\theta,\lambda)$ from the aqueous suspensions of the
microspheres with the diameter $d$ = 1.5 $\mu m$ (standard
deviation, S.D. = 0.04 $\mu m$) ($g$ = 0.93 at $\lambda =
550~nm$). We varied $l_s$ from 3 $\mu m$ to 800 $\mu m$ with the
fixed $L_{sc}$ ($L_{sc} = 35~\mu m$). As $l_s$ increases, the Mie
scattering features such as the oscillatory pattern and the slope
of the overall decline of intensity with wavelength become obvious
and prominent. These spectral features indicate that only a few
scattering events give rise to EBS. Increasing $l_s$ reduces the
number of particles in the coherence volume that is determined by
$L_{sc}$, hence only lower orders of scattering events contribute
to the EBS peak. For example, the spectrum of the sample with $l_s
= 3~\mu m$ resembles that of the diffuse multiple scattering of
highly packed media. As $l_s$ increases, the Mie scattering
patterns are revealed. Finally, for $l_s\!>>\!L_{sc}$, the
spectral shape remains unchanged, indicating that the parametric
condition reaches to the minimal number of scattering events
required for EBS.

To explore quantitatively the observations above, we developed a
Mie theory-based double scattering model, which provides the
backscattering spectrum and the angular profile of EBS from double
scattering. The radial intensity probability $P(r,\lambda)$ of
double scattering can be expressed as
\begin{eqnarray}
P(r,\lambda)=\int^\infty_0 \!\!\!\! \int^\infty_0 \!\!
\frac{\exp(-\mu
_a(\sqrt{r^2+(z-z')^2}+z+z'))}{[r^2+(z-z')^2](z'+d)^2}
\nonumber \\
\times~\mu_s(\lambda)F(\Theta,\lambda)\mu_s(\lambda)F(\pi-\Theta,\lambda)dzdz',
\end{eqnarray}
where $r$ is the transverse radial distance between two
scatterers, $z$ and $z'$ are the vertical distances from the
surface to the scatterers, respectively, $\Theta
=\tan^{-1}(r/(z-z'))$, $F(\Theta)$ is the phase function of single
scattering, $d$ is the diameter of the microscophere, and $\mu_a$
is an attenuation coefficient.  $\mu_a$ was obtained separately by
measuring reflectance intensity for various sample thicknesses and
calculating the exponential attenuation length. $F(\Theta)$ and
$\mu_s$($=1/l_{s}$) were calculated using Mie
theory~\cite{van_de_Hulst:1995}.

\begin{figure}
\includegraphics[scale=0.80]{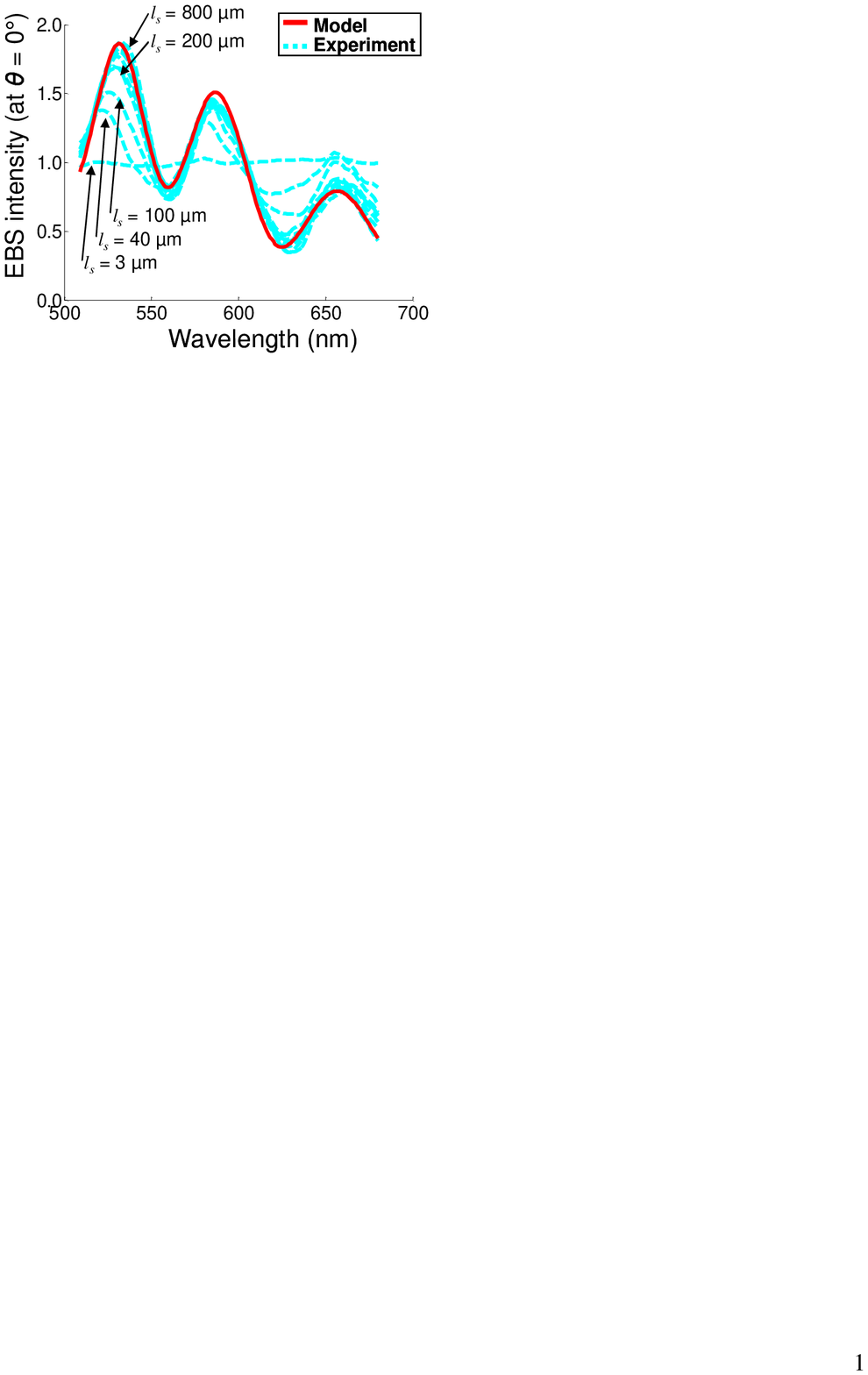}
\caption{\label{fig:fig2} The predictions of the double scattering
model in the spectral features of LEBS in the backward direction.
}
\end{figure}
\begin{figure}
\includegraphics[scale=0.95]{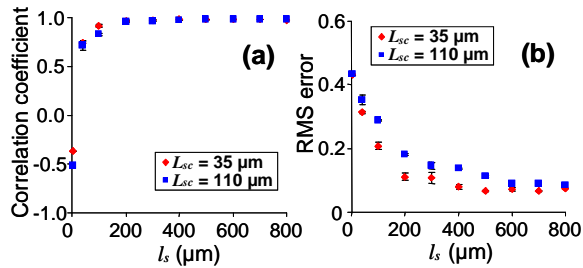}
\caption{\label{fig:fig3} Accuracy analyses of the double
scattering model of LEBS.  (a) Correlation coefficient.  (b) RMS
error. }
\end{figure}
We also obtained the angular profile of the EBS peak from
$P(r,\lambda)$ of double scattering. $I_{EBS}(\theta,\lambda)$ can
be expressed as an integral transform of the path length
distribution of the conjugated time-reversed light
paths~\cite{Akkermans:PRL:1986}:
$I_{EBS}(q_{\bot})=\int\!\!\!\int
C(r)P(r)\exp(i\textbf{\textit{q}}_{\bot}\!\cdot\textbf{\textit{r}})d^2r,$
where $\textbf{\textit{q}}_\bot$ is the projection of the wave
vector onto the plane orthogonal to the backward direction,
$P(\textbf{\textit{r}})$ is the probability of the radial
intensity distribution of EBS photons with the radial vector
$\textbf{\textit{r}}$ pointing from the first to the last points
on a conjugated time-reversed light path ($\textbf{\textit{r}}$ is
perpendicular to the incident light), and $C(r)
=|2J_1(r/L_{sc})/(r/L_{sc})|$ is the degree of spatial coherence
with $J_1$ the first order Bessel function~\cite{Bron:1999}. If a
medium is isotropic, the two-dimensional Fourier integral becomes
the Fourier transform of $C(r)rP(r)$:
\begin{equation}
I_{EBS}(\theta)\propto \int^\infty_0 C(r)rP(r)\exp(i2\pi r \theta
/ \lambda)dr,
\end{equation}
with $q_\bot  =2\pi \theta /\lambda$.  As a result, the width of
the LEBS peak is inversely proportional to $C(r)rP(r)$.

Figure 2 shows that for $l_s$ greater than a few $L_{sc}$, the
spectral shape of $I_{EBS}(\theta=0^\circ,\lambda)$  approaches
the spectrum of the double scattering model and then remains
unchanged for $l_s\!>>\!L_{sc}$.  In order to quantify the
agreement between the model and the experimental results, we used
two complimentary measures: the root mean square (RMS) error and
the correlation coefficient. The RMS error measures the overall
estimation accuracy while the correlation coefficient measures the
capability of the double scattering model to replicate the
oscillation characteristics of the experimental spectra.  Figure 3
plots the correlation coefficient (Fig. 3(a)) and the RMS error
(Fig. 3(b)) as a function of $l_s$ for two different values of
$L_{sc}$ ($L_{sc} = 35~\mu m$ and $L_{sc} = 110~\mu m$).  In both
cases, as shown in Fig. 3, the two measures level off for $l_s\!>
\!4L_{sc}$, thus indicating that the double scattering model is in
excellent agreement with the experimental data.

Figure 4 compares $I_{EBS}(\theta,\lambda)$ obtained
experimentally with the predictions of the double scattering model
for two different values of $L_{sc}$.  We convoluted the profiles
of the LEBS peaks with the angular response of the instrument to
take into account the finite point-spread function of the
detection system and the incident beam divergence.  As can be seen
from Fig. 4, the double scattering model is in excellent agreement
with experimental data and predicts both the angular and spectral
profiles of LEBS. As expected from the Fourier transform
relationship between $I_{EBS}(\theta)$ and $C(r)rP(r)$ (Eq.~(2)),
the shorter $L_{sc}$ generates the broader LEBS peak as shown in
Fig. 4. These results confirm the hypothesis that in low-coherence
regime ($L_{sc}\!<<\!l_s$) the number of scattering events giving
rise to EBS reaches its minimum and LEBS is indeed generated by
means of double scattering.


\begin{figure}
\includegraphics[scale=0.95]{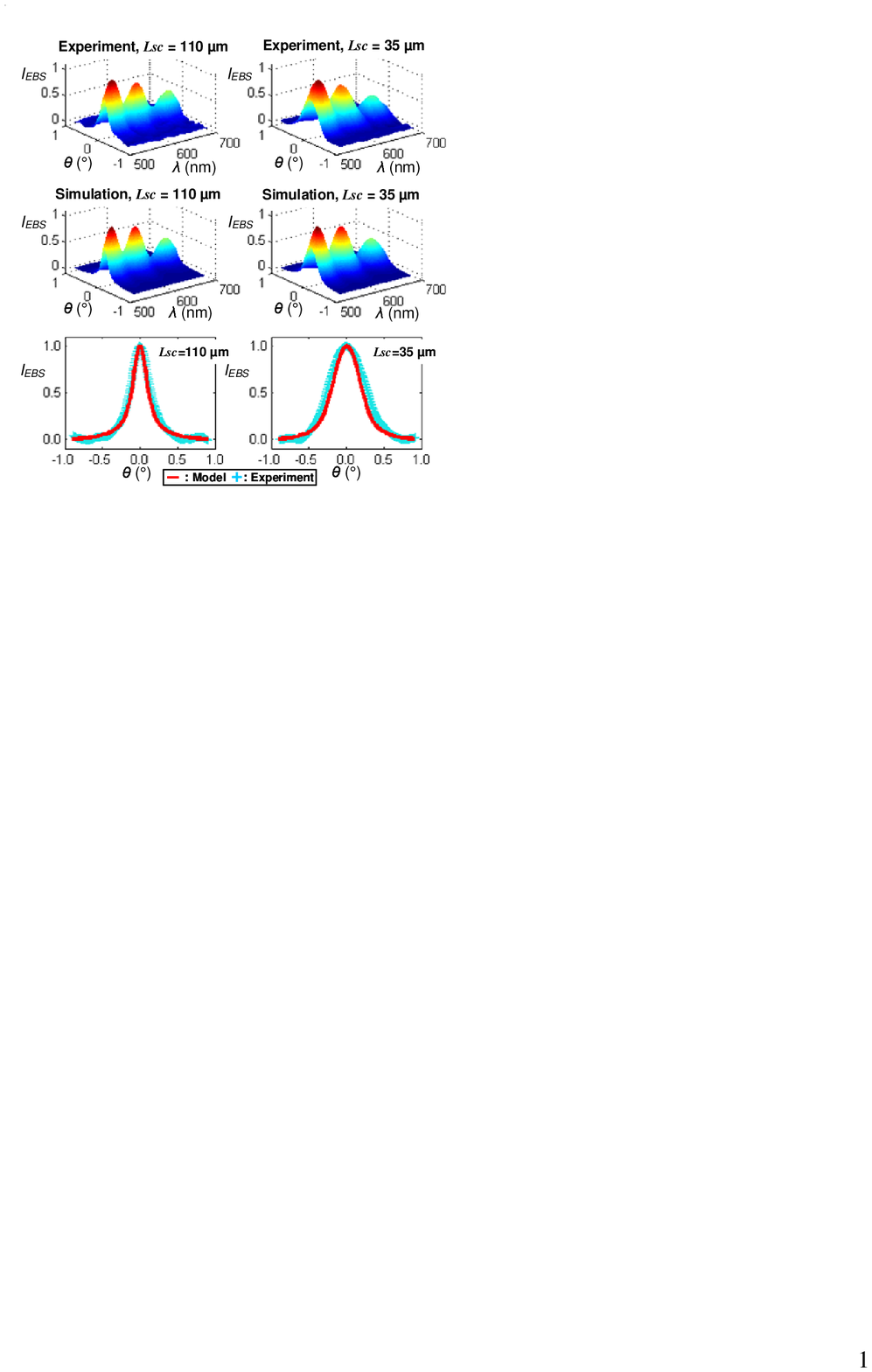}
\caption{\label{fig:fig4} $I_{EBS}(\theta,\lambda)$ obtained from
the experiment of the aqueous suspension of microspheres ($d$ =
1.5 $\mu m$) and its simulation from the double scattering model
for $L_{sc} = 35~\mu m$ and 110 $\mu m$ with $l_s$ = 700 $\mu m$.
The angular profiles of the LEBS peaks at $\lambda$ = 532 $nm$.}
\end{figure}

In LEBS, \textit{a priori} surprisingly, the LEBS peaks from the
helicity preserving $(cir||cir)$ channel are lower than those from
the orthogonal helicity $(cir\bot cir)$ channel. Conventionally,
the EBS peaks from the $(cir||cir)$ channel are much higher than
those from the
 $(cir\bot cir)$ channel. Figure 5 shows the
enhancement factor $(\equiv I(\theta=0^\circ)/I_{BASE})$ of LEBS
peaks at $\lambda = 520~ nm$ recorded from a discrete random
medium consisting of aqueous suspension of the microspheres ($d =
0.20~\mu m$, S.D. = 0.01 $\mu m$) from the both channels with
$L_{sc}= 35~\mu m$. From Mie theory,
the forward scattering preserves the degree of circular
polarization, while the backscattering flips the circular
polarization in a manner similar to light reflection from a
mirror.
In LEBS, the direction of light scattered by one of the scatterers
should be close to the forward direction while the direction of
light scattered by the other scatterer should be close to the
backward direction. Therefore, the enhancement factor from the
$(cir\bot cir)$ channel is consistently higher than that from the
$(cir||cir)$ channel, and the difference in the enhancement
factors between the $(cir\bot cir)$ channel and the $(cir||cir)$
channel is nearly constant for $l_s\!>>\!L_{sc}$, supporting the
validity of the double scattering model of LEBS.

%

\begin{figure}
\includegraphics[scale=0.65]{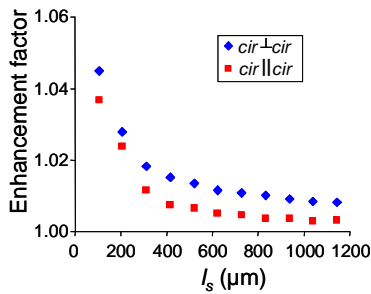}
\caption{\label{fig:fig5} Enhancement factor of LEBS in the
circular polarization channels for $L_{sc}$ = 35 $\mu m$.  The
enhancement factor from the $(cir\bot cir)$ channel  is higher
than that from the $(cir||cir)$ channel. }
\end{figure}
In conclusion, (i) using low coherence illumination, we were able
to create a virtual coherence volume within random media, which
rejects longer paths and isolates lower order scatterings in EBS.
(ii) Controlling the coherence length of illumination and the
optical properties of the discrete random media, we were able to
isolate double scattering in EBS.  This led for the first time to
prove an existing theoretical hypothesis that the minimum number
of scattering events needed to generate EBS is double scattering.
(iii) We demonstrated that a large number of the independent
coherence areas provide statistical information about the optical
properties of random media without configuration or ensemble
averaging.  (iv) We reported experimental results for $L_{sc} <<
l_s^*$ in weakly scattering disordered media.  In this
dramatically different regime, $L_{sc}$ can be made to be the
shortest length scale except the particle size.  (v) In the
previous publications ~\cite{Kim:OL:2004,Kim:OL:2005}, we showed
that LEBS signals from human colon are sensitive to early
precancerous alterations in colon cancer.  However, the origin of
LEBS in colonic mucosa has not been completely understood.
Therefore, our finding that EBS originates from the time-reversed
paths of double scattering events in weakly scattering media will
further facilitate understanding of EBS signals for tissue
diagnosis and characterization, providing a potential method about
how to analyze the LEBS signals from biological tissue.

Correspondence to: \\Young L. Kim at younglae@northwestern.edu

\end{document}